\documentclass[prl,twocolumn,superscriptaddress,notitlepage,9py]{revtex4-1}
\usepackage[utf8]{inputenc}
\usepackage{natbib}
\usepackage{graphicx}
\usepackage{color}
\usepackage[left=1.5cm, right=1.5cm, top=1.785cm, bottom=2.0cm]{geometry}
\usepackage{amsmath}
\usepackage{floatrow}

\begin{document}

\title{\LARGE Scaling of relaxation and excess entropy in plastically deformed amorphous solids}

\author{{\Large K. Lawrence Galloway$^{1,*}$, Xiaoguang Ma $^{3,a,*}$, Nathan C. Keim$^{4}$, Douglas J. Jerolmack$^{5,1}$, Arjun G. Yodh$^{3}$, Paulo E. Arratia$^{*,1}$ \\}
{\large \small $^{1}$Department of Mechanical Engineering and Applied Mechanics, University of Pennsylvania \\ $^{2}$Department of Bioengineering, University of Pennsylvania \\ 
$^{3}$Department of Physics and Astronomy, University of Pennsylvania \\ 
$^{4}$Department of Physics, Pennsylvania State University \\ 
$^{5}$Department of Earth and Environmental Science, University of Pennsylvania \\
$a$ To whom correspondence may be addressed. Email: xiaom@seas.upenn.edu \\
$*$ These two authors contributed equally. \\
}}

\date{\today}

\begin{abstract}
    \normalsize{\textbf{ When stressed sufficiently, solid materials yield and deform plastically via reorganization of microscopic constituents. Indeed, it is possible  to alter the micro-structure of materials  by judicious application of stress, an empirical process utilized in practice to enhance the mechanical properties of metals. Understanding the interdependence of plastic flow and microscopic structure in these non-equilibrium states, however, remains a major challenge. Here, we experimentally investigate this relationship, between the relaxation dynamics and microscopic structure of disordered colloidal solids during plastic deformation. We apply oscillatory shear to solid colloidal monolayers and study their particle trajectories as a function of shear rate in the plastic regime. Under these circumstances, the strain rate, the relaxation rate associated with plastic flow, and the sample microscopic structure oscillate together but with different phases. Interestingly, the experiments reveal that the relaxation rate associated with plastic flow at time $t$ is correlated with the strain rate and sample microscopic structure measured at earlier and later times, respectively. The relaxation rate, in this non-stationary condition, exhibits power-law shear-thinning behavior and scales exponentially with sample excess entropy. Thus, measurement of sample static structure (excess entropy) provides insight about both strain-rate and constituent rearrangement dynamics in the sample at earlier times.}}
\end{abstract}

\maketitle

\section{Introduction}

For many amorphous solids, i.e., solids without long-range order, a threshold stress exists beyond which the material starts to deform plastically (yield) and flow like a liquid. These yield stress materials, which range from foams and colloids to cement and metallic glasses, have constituents and dynamics that vary widely across length and time scales \cite{doi:10.1146/annurev-fluid-010313-141424,RevModPhys.89.035005,RevModPhys.90.045006}. Nevertheless, they are unified by two features: the cross-over transition from solid- to liquid-like behavior and a nonlinear viscosity response to external stress (shear thinning) \cite{Seth2011}. Ultimately, to understand these nonlinear mechanical processes, we need a detailed picture about how shear couples to microscopic structure and relaxation. If successful, this understanding could lead to improved processing of amorphous metals via stress-induced control of microstructure \cite{ENGLER2002249,Sun2016}.

To this end, useful models have been developed to characterize the structural origin of plasticity in amorphous solids. Shear transformation zone models, for example, posit the existence of mechanically weak regions in amorphous solids analogous to crystalline defects, and then they focus (largely) on the kinetics associated with localized plastic events \cite{PhysRevE.57.7192}. The softer regions are believed to facilitate or accelerate rearrangements nearby. This general phenomenology of dynamic heterogeneity is observed in experiments \cite{Schall1895} and computer simulations \cite{PhysRevE.57.7192}, and is supported by first-principle Mode-Coupling and Random First-Order Transition theories \cite{Gotze2008,PhysRevA.40.1045,Lubchenko11506}. Nevertheless, identification of mechanically weak regions from {\it static} sample structure, e.g., before plastic events occur, remains a challenge.

In a different vein, thermodynamic predictors based on excess entropy ($S^{ex}$) have shown promise for explaining nonlinear mechanical phenomena in complex fluids \cite{Rosenfeld77,Dzugutov96,PhysRevE.78.010201,Abramson2009,Tanaka18,Dyre2018}. Excess entropy concepts developed from studies of liquids rather than solids and enable comparison of macroscopic system-averaged structural and dynamical quantities. $S^{ex}$ is a structural order parameter defined as the difference between system thermodynamic entropy and that of an equivalent ideal gas \cite{Baranya89}. For typical liquids, $S^{ex}$ derives mainly from pair correlations of its constituents \cite{Dzugutov96} and is readily evaluated by experiment \cite{Abramson2009,Ma2013}. Excess entropy accurately predicts transport coefficients of simple and complex fluids in equilibrium using their static structure \cite{PhysRevLett.92.145901,Hoyt00,Samanta01,Truskett06,Krekelberg2009,Abramson2009,doi:10.1063/1.3414349,Truskett11,Ma2013,Chen2015,PhysRevLett.122.178002,doi:10.1063/1.5091564,Li2018}. Recently, in computer simulations, $S^{ex}$ has been applied to supercooled liquids under steady-state shear; the shear-dependent relaxation time of the supercooled liquids was found to scale with $S^{ex}$ \cite{Tanaka18}, thereby revealing a simple structural connection to shear-thinning induced relaxation. This intriguing discovery has not been tested experimentally. Moreover, the concept of excess entropy scaling has not been applied to understand plastic flow in amorphous solids, nor in materials driven into more general non-stationary states.

\begin{figure*}
\centering
\includegraphics[width=1.\linewidth]{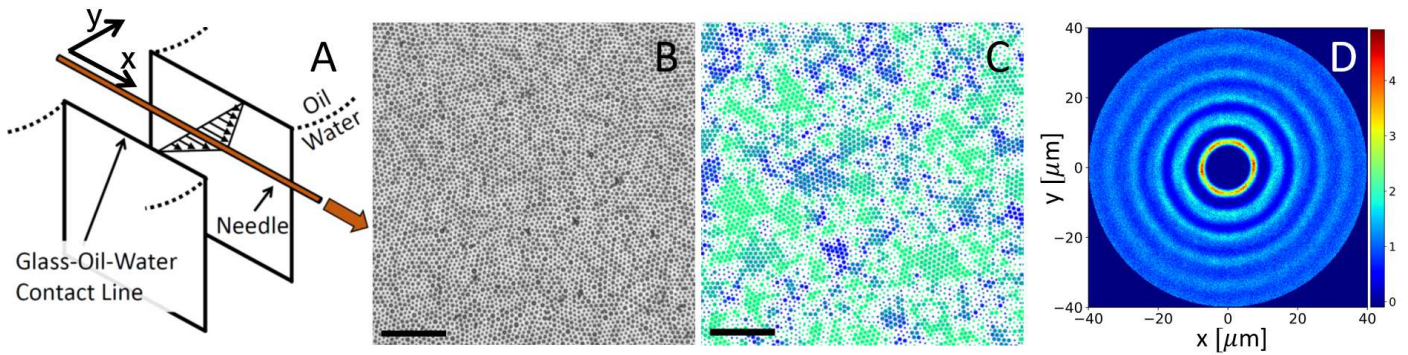}
\caption{(a) Schematic of the interfacial stress rheometer. A sinusoidal magnetic force is imparted to the interface-bound magnetic needle, which in turn introduces oscillatory shear stress at the oil-water interface. The parallel and perpendicular directions with respect to the needle motion are defined as the $x$- and $y$-axis, respectively. (b) Micrograph of bi-disperse colloidal particles at the oil-water interface from sample A. (c) Sixfold bond orientation order, $\psi_6$, measured from particles in (b). Colors help to indicate the lattice director (orientation) as a guide for the eye to help discern ordered and disordered domains. Dots with large size indicates $|\psi_6|>0.9$, and small dot size indicates $|\psi_6|<0.9$. The scale bars in (b) and (c) are $100\mu$m. (d) A pair correlation function, $g(x,y)$, measured from particle positions in (b) exhibits strong anisotropy due to ordered domains.}
\label{fig1}
\end{figure*}

In this contribution, we investigate the connection between shear rate, relaxation time and excess entropy of plastically deformed matter in non-stationary states. We use a custom-made interfacial stress rheometer \cite{Shahin86,Reyn08,C3SM51014J,PhysRevLett.112.028302} to apply oscillatory shear at different strain amplitudes to an oil-water interface (Fig.~\ref{fig1}a, see {\it Materials and Methods} for details). A series of disordered, two-dimensional colloidal solids are prepared at the oil-water interface (Fig.~\ref{fig1}b). Their translational and orientational correlation functions do not exhibit long-range order (see the Supporting Information (SI)). The disordered samples are driven by the applied oscillatory shear, and concurrently, the trajectories of individual particles in the samples are captured by video optical microscopy and standard tracking software.

From particle position data during oscillatory shear, we compute strain rate, the relaxation rate/time associated with plastic flow, and the sample excess entropy. The relaxation time exhibits a power-law scaling with shear rate, a characteristic of shear-thinning behavior. Furthermore, phase-shifts between the oscillatory signals revealed a constant lag time between plastic shear rate and plastically induced relaxation rate, and a different lag time (proportional to the instantaneous relaxation time) between relaxation rate and excess entropy. These delay-intervals (phase-shifts) uncover novel connections between shear rate, plastic flow induced relaxation, and structure of the samples in non-stationary states. Surprisingly, we find that relaxation time/rate and excess entropy data measured at different strain amplitudes collapse onto a single master exponential scaling curve which depends only on sample type. In total, the work introduces an analysis framework based on excess entropy scaling to understand plastic flow in both stationary and non-stationary states, and the findings suggest that information about the relaxation history of an amorphous material can be deduced from its current static structure.

\begin{table}
\small
  \label{tab:table1}
  \begin{tabular*}{1.0\textwidth}{@{\extracolsep{\fill}}lllllll}
    \hline
    Sample & $\sigma$ ($\mu\mathrm{m}$) & $\phi$ & d ($\mu\mathrm{m}$) & $\Gamma_{\mathrm{max}}$ \\
    \hline
    A (bi-disperse)  & 4.1,5.6 & 43\% & 7.4 & 5-16\%\\
    B (monodisperse) & 5.6     & 32\% & 7.7 & 8-16\%\\
    C (bi-disperse)  & 1.0,1.2 & 32\% & 9.8 & 5-8\%\\
    \hline
  \end{tabular*}
  \caption{\ Summary of colloidal monolayers. $\sigma$: particle diameter, $\phi$: packing fraction, d: mean interparticle separation, $\Gamma_{\mathrm{max}}$: strain amplitude }
\end{table}

Briefly, the solid-like monolayers consist of colloidal spheres with different diameters ($\sigma$), surface charge densities, and packing fractions ($\phi$) (see Table.1 and {\it Materials and Methods}). In combination, these factors determine interparticle separation (d), sample structure (see Fig.~\ref{fig1}c,d and SI), shear moduli, and plasticity \cite{C3SM51014J,PhysRevLett.112.028302}. Rheology measurements of the samples exhibit elastic behavior at small strain amplitudes and yielding behavior when the strain exceeds about $3\%$ (see SI). Herein, we focus exclusively on strain amplitudes above yield point (e.g., larger than $5\%$, see Table 1).

\begin{figure*}
\centering
\includegraphics[width=0.77\linewidth]{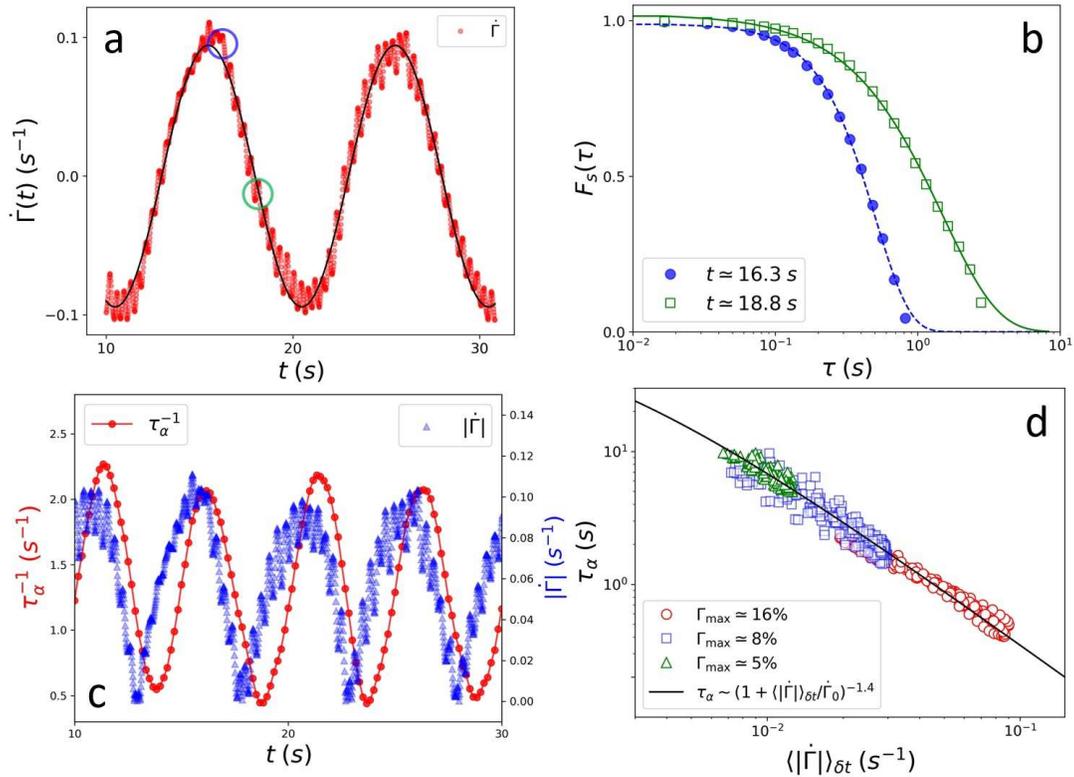}
\caption{Dynamics in sample A (a) Instantaneous shear rate, $\dot{\Gamma}(t)$, versus time, $t$. The solid line is the sinusoidal fit, $\dot{\Gamma}(t)=0.096\mathrm{sin(\omega t)}$. (b) The self-part of the intermediate scattering function, $F_s(\tau)$, measured at the two times indicated by same-color circles (green, blue) in (a). The dashed and solid lines are fits using Eq.~\ref{eq_exponential}, with $\tau_\alpha=0.5\;\text{and}\;2.3$ sec and $\beta=1.3\;\text{and}\;1.5$ at $t=16.3\;\text{and}\;18.8$ sec, respectively. (c) Relaxation rate, $\tau_\alpha^{-1}(t)$, versus time, $t$ (red circles). The magnitude of shear rate, $|\dot{\Gamma}(t)|$, is also plotted (blue triangles) for phase-shift comparison. (d), The measured relaxation time, $\tau_\alpha(t+\Delta t)$, versus time-averaged shear rate, $\langle|\dot{\Gamma}(t)|\rangle_{\delta t}$, from three experiments with different $\Gamma_{\text{max}}$ values. The solid line is the best fit using $\tau_\alpha\sim (1+\langle|\dot{\Gamma}|\rangle_{\delta t}/0.0017)^{-1.4}$.}
\label{fig2}
\end{figure*}

We first use the particle trajectory data to measure and compare shear rates and shear-induced relaxation times. The shear strain, $\Gamma(t)$, at time $t$ quantifies the sample's {\it affine} deformation, which follows the oscillations of the needle motion. We compute $\Gamma(t)$ by taking average of the measured $y$-dependent $\it local$ strain, $\gamma(y,t)$ (see {\it Materials and Methods}). Figure~\ref{fig2}a shows that $\dot{\Gamma}(t)$ follows the driving sinusoidal function set by the external force, and that it exhibits measurable fluctuations about the sinusoidal function too. Note, fluctuations of $\Gamma(t)$ about the driving stress have been seen in plastically deformed bidisperse  polycrystals in computer simulations \cite{PhysRevE.77.042501,PhysRevE.81.051501}; these fluctuations were attributed to intermittent yielding  along grain boundaries and become weaker when the sample has smaller crystalline domains. 

We use {\it nonaffine} particle motions to evaluate sample relaxation behavior \cite{PhysRevE.58.3515,Tanaka18}. At time $t$, the self-part of the intermediate scattering function is,
\begin{equation}
F_s(t,\tau)=\dfrac{1}{N}\langle\sum_{j=1}^{N}\mathrm{exp}[\frac{2\pi i}{d}|\Delta\vec{r}\:'_j(\tau)|]\rangle.
\label{eq_scattering}
\end{equation}
Here $N$ is the number of particles and $\Delta\vec{r}\:'_j$ the nonaffine displacement of the $j$-th particle, that is, the residual after the affine displacement has been subtracted from the total particle displacement, $\Delta\vec{r}_j$. (See {\it Materials and Methods} and SI for how to compute $\Delta\vec{r}\:'_j$ from $\Delta\vec{r}_j$.) The brackets, $\langle\cdots\rangle$, represent a time average over the period $[t-\delta t/2, t+\delta t/2]$ ($\delta t=2.5$s is one quarter of the shear cycle). The duration of the measurement is thus $\delta t$. Ideally, $F_s(\tau)$ should decay to below $1/e$ at $\tau=\delta t$ to extract the relaxation time. However, we will soon show this is not necessary.

Figure~\ref{fig2}b shows examples of $F_s(t,\tau)$ at two times where the $\dot{\Gamma}$ values are different; these $F_s(t,\tau)$'s decay at different rates, indicating shear-dependent relaxation behavior. $F_s(t,\tau)$ is well fit by the function,
\begin{equation}
F_s(t, \tau)=A\mathrm{exp}[-(\tau/\tau_\alpha)^\beta],
\label{eq_exponential}
\end{equation}
where $\tau_\alpha$ is the $\alpha$-relaxation time measured in the time-interval centered on $t$, and $A\simeq 1$ is a constant prefactor (see SI). Since we can fit $F_s(\tau)$ data before it decays to $1/e$ to obtain $\tau_\alpha$, we can estimate $\tau_\alpha$ from measurements with duration ($\delta t$) shorter than $\tau_\alpha$ (see SI for details). Interestingly, we find $\beta>1$ (compressed exponential) throughout the shear cycle in all samples. This finding confirms the expectation that particle configurations, when driven by external forces, relax/reorganize faster than would occur if driven by exponential diffusive motions alone. A few studies have also reported $\beta>1$ phenomena \cite{PhysRevLett.84.2275,Masri_2005,Madsen_2010,PhysRevE.79.011501,Angelini2014,Gnan2019}; in these cases, ballistic motions of constituents were found to accompany the accumulation and release of internal stress \cite{PhysRevLett.113.078301}. In our experiments the nonaffine mean-square-displacements (MSD's), $\langle\Delta r'^2(\tau)\rangle$, exhibit super-diffusive behavior, that is, $\langle\Delta r'^2(\tau)\rangle\sim\tau^p$ with $p>1$ (see SI); by analogy to prior work, we believe the measured compressed exponential decay of $F_s(t,\tau)$ is caused by super-diffusive particle motions. Note, we also investigated other alternative explanations for the compressed exponential decay of $F_s(\tau)$ (see SI).

Using the $\dot{\Gamma}(t)$ and $\tau_\alpha(t)$ data, we next investigate how shear influences relaxation in the {\it non-stationary} regime. Figure~\ref{fig2}c compares $|\dot{\Gamma}(t)|$ and $\tau_\alpha^{-1}(t)$ measured from sample A ($\Gamma_{\text{max}}=16\%$) as a function of $t$; here, the absolute shear rate is used because we expect the shear direction to have little influence on relaxation rate. This comparison clearly demonstrates that the relaxation rate lags the shear rate by a time interval, $\Delta t\simeq 0.8$s (see {\it Materials and Methods} and SI). This lag time hints at a causal relation between shear and shear-induced relaxation processes. Moreover, the amplitude of $\tau_\alpha^{-1}(t)$ follows $|\dot{\Gamma}(t)|$. 

For a more quantitative comparison, we examine our data in the context of the non-Newtonian relationship between shear rate and relaxation time that has been found in steady-state \cite{doi:10.1063/1.331272,PhysRevE.61.5464}:
\begin{equation}
\tau_\alpha\sim (1+\dot{\Gamma}/\dot{\Gamma}_0)^\mu.
\label{eq_non}
\end{equation}   
Here, $\dot{\Gamma}_0$ is the shear rate associated with onset of non-Newtonian viscous response behavior, and $\mu<0$ is a power law exponent characterizing shear-thinning behavior. 

In oscillatory measurements, Eq.~\ref{eq_non} has been established between the {\it mean} (or {\it maximal}) viscosity and shear rate during multiple shear cycles \cite{Cheng1276}. To our knowledge, this relation has not been used to describe the connection between the {\it instantaneous} viscosity (or relaxation time) and shear rate in non-stationary samples. Despite the phase-shift between shear rate and shear-induced relaxation rate, we might expect our data to follow Eq.~\ref{eq_non} with $\dot{\Gamma}$ being replaced by its weighted time average, $\langle|\dot{\Gamma}|\rangle_{\delta t}$, over the time-interval $[t-\delta t/2, t+\delta t/2]$ (that is, the same window wherein $\tau_\alpha(t)$ is evaluated, see {\it Materials and Methods} and SI about calculation of $\langle|\dot{\Gamma}|\rangle_{\delta t}$). To test this hypothesis, we plot $\tau_\alpha(t+\Delta t)$ versus $\langle|\dot{\Gamma}(t)|\rangle_{\delta t}$ measured from sample A in Fig.~\ref{fig2}d. Remarkably, the data from the sample A sheared at three different strain amplitudes collapse onto a single master curve; the best fit using Eq.~\ref{eq_non} gives $\dot{\Gamma}_0=(1.7\pm 0.5)\times 10^{-3}\;s^{-1}$ and $\mu=-1.4\pm 0.3$. Interestingly, the fitted $\dot{\Gamma}_0$ in our sample is of the same order of magnitude as the onset shear rates of non-linear viscous response in molecular glasses \cite{doi:10.1063/1.331272}. The fitted $\mu$ is similar to those measured in dense suspensions of soft colloidal particles \cite{rug01:002006527}. This finding suggests an interesting new way to characterize shear-thinning behavior in a non-stationary (e.g., oscillatory) measurement. Note, also, while in principle the lag time, $\Delta t$, between shear-rate and relaxation time may be a complex function of shear rate, in our samples it suffices to use a constant lag time.

Next, we compute excess entropy from particle position data and explore whether excess entropy scaling laws can be applied in systems experiencing {\it non-stationary} (oscillatory) shear. If the scaling relation still holds, then by implication, sample static structure can provide information about relaxation induced by plastic deformation. Previously, viscosity, diffusion coefficients, and relaxation times have been found to obey a simple excess entropy scaling law, $\tau_\alpha\sim f(S^{ex})$, for a wide variety of materials \cite{ PhysRevLett.92.145901,Hoyt00,Samanta01,Truskett06,Krekelberg2009,doi:10.1063/1.3414349,Truskett11,Chen2015,PhysRevLett.122.178002,doi:10.1063/1.5091564,Li2018}  spanning different particle type, size, density, interaction, temperature, material phase, and even shear rate \cite{Krekelberg2009,Tanaka18}. Importantly, $S^{ex}$ is well approximated by the two-body contribution, $S_2$, which is readily determined from scattering or imaging experiments \cite{Abramson2009,Ma2013,Chen2015}.

\begin{figure*}
\centering
\includegraphics[width=0.78\linewidth]{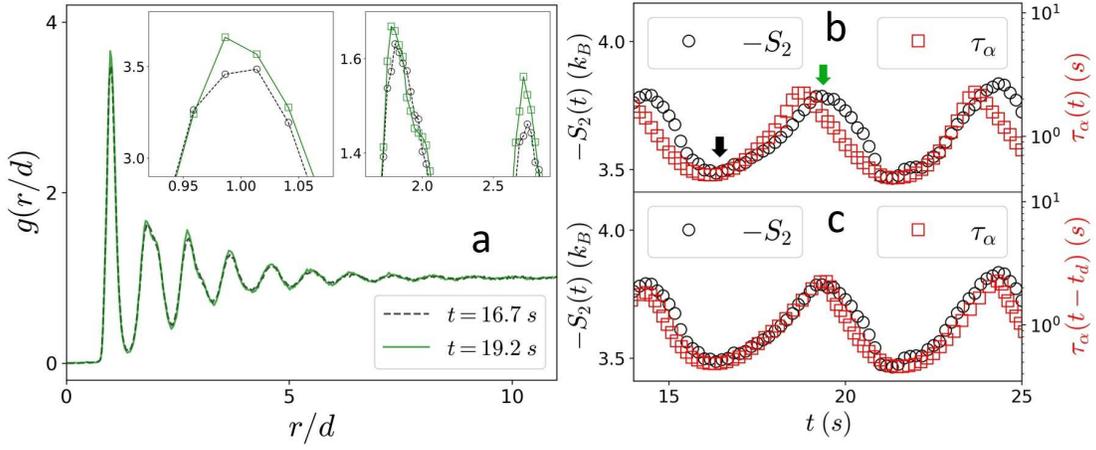}
\caption{(a) Measured $g(r)$ from data taken at $t=16.7\;\text{and}\;19.2\;s$. The insets show the enlarged plots of the first (left inset), and the second and third peaks (right inset) of $g(r)$, respectively. (b) $S_2(t)$ versus $t$. $\tau_\alpha(t)$ is also plotted for comparison. The black and green arrows indicate $t=16.7\;\text{and}\;19.2\;s$, respectively. (c) $\tau_\alpha(t-t_d)$ is plotted versus time delay, $t_d(t)\simeq 0.3\tau_\alpha(t)$.}
\label{fig3}
\end{figure*}

To this end, we compute the time-dependent sample pair correlation function, $g(r)$, using particle positions at time $t$. Note, we employ particle coordinates in a single video frame at time $t$ for determination of $g(r)$; these particles are the same as used above in computing $\dot{\Gamma}(t)$ and $\tau_\alpha(t)$. Examples of $g(r)$ at two times, $t=16.7\;\text{and}\;19.2\;s$, are shown in Fig.~\ref{fig3}a. These $g(r)$’s exhibit quasi-long-range order extending out to 10 shells of neighbors; the extended correlations are indicative of presence of many small crystalline domains (see Fig.~\ref{fig1}c and SI). By comparison, $g(r)$ from sheared glass-forming liquids typically exhibits only 3 well-defined peaks (e.g., see Refs.~\cite{PhysRevE.58.3515,Tanaka18}). The correlation lengths obtained from the spatial correlations of translational and orientational order also confirmed that the samples are more ordered than traditional glasses but less ordered than crystals/polycrystals (see SI).

The peaks of $g(r)$ evolve subtly throughout imposed shear cycles (Fig.~\ref{fig3}a insets); these changes are indicative of shear-induced restructuring. The comparatively high peaks in $g(r)$ at $t=19.2$ s compared to $t=16.7$ s suggests a more ordered structure in the former case. The differences in peak height at different times are rather small and are in accord with measurements in sheared molecular glasses \cite{PhysRevE.58.3515,PhysRevE.78.010201,Tanaka18}. From the time-dependent $g(r)$ data, we compute $S_2$ versus $t$,
\begin{equation}
S_2=-\pi \rho\int_0^{\infty}\{g(r)\ln[g(r)]-[g(r)-1]\}rdr.
\label{eq_entropy}
\end{equation}
Here, $\rho$ is sample particle number density. Equation~\ref{eq_entropy} converges quickly after $r$ reaches $5d$ (see SI); thus we choose the cut-off length, $ r_{\mathrm{cut}}=10d$, as the integration limit for computing $S_2$. We confirmed that with the same cut-off length Eq.~\ref{eq_entropy} converges for the other two samples as well (see SI).A larger $-S_2$ value at $t=19.2$ s confirms a more ordered structure, consistent with $g(r)$ data in Fig.~\ref{fig3}a. Note that larger $-S_2$ are also accompanied by larger bond orientation order \cite{Tanaka2010}, suggesting the orientational order is coupled to translational order by shear (see SI).

Figure~\ref{fig3}b presents $S_2(t)$ and $\tau_\alpha(t)$ as a function of $t$ during the shear cycles. Notice that longer relaxation times, $\tau_\alpha$, are accompanied by larger $-S_2$ or, equivalently, more ordered sample structures. Taken together with the $\dot{\Gamma}(t)$ findings (Fig.~\ref{fig2}c), we conclude that faster shear rates lead to shorter shear-induced relaxation times and more disordered resultant particle arrangements. 

Further inspection of Fig.~\ref{fig3}b reveals that the peaks of $-S_2(t)$ clearly lag behind those in $\tau_\alpha(t)$; by comparison, the lags between valleys are less apparent. A possible explanation is that $S_2(t)$ lags behind $\tau_\alpha(t)$ at all phase positions; this hypothesis is further supported by the hysteresis loops generated by the two functions (see SI). Based on this intriguing observation, we hypothesize that: i) the relaxation time measured at $t$ is related to the sample structure ($S_2$) at a later time, $t+t_d$, and ii) the time delay, $t_d(t)$, is a function of $\tau_\alpha(t)$. We assume $t_d(t)\simeq h\tau_\alpha(t)$, where $h$ is a constant throughout the shear cycles. To test this hypothesis we re-plot $\tau_\alpha(t-t_d)$ in Fig.~\ref{fig3}c. The relation $t_d(t)\simeq 0.3\tau_\alpha(t)$ best aligns the peaks and the valleys of $\tau_\alpha(t-t_d)$ and $-S_2(t)$ (see {\it Materials and Methods} and SI). Note, the choice of a linear function of $\tau_\alpha$ to approximate $t_d$ is empirical; $t_d$ could have a more complex dependence on $\tau_\alpha$, $\dot{\Gamma}$, and their time-derivatives. This empirical finding that $t_d\sim\tau_\alpha$ suggests a picture wherein new structures driven by shear-induced relaxation evolve to their final form after a waiting time that is itself dependent on the relaxation process/timescale. Moreover, the introduction of this form for $t_d$ enables comparison of $\tau_\alpha$ versus $S_2$ across different times and conditions. 
 
\begin{figure}[hb!]
\centering
\includegraphics[width=1.\linewidth]{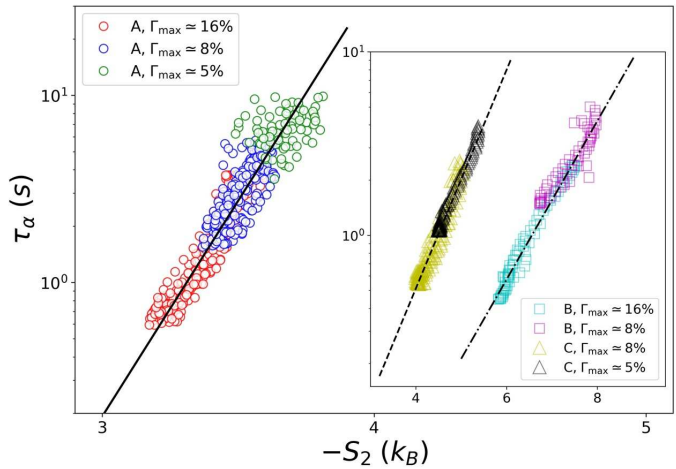}
\caption{Delayed relaxation time, $\tau_\alpha(t-t_d)$, as a function of the excess entropy, $-S_2(t)$, measured from sample A with 3 different $\Gamma_{\text{max}}$. Inset shows same data measured from sample B and C. The three data sets are fit using Eq.~\ref{eq_scaling} with $c=3.9\pm 0.2 (\text{solid line}),\;1.0\pm 0.1 (inset \text{dash-dot line}),\;\text{and}\;1.4\pm 0.1 (inset \text{dashed line})$ for sample A, B, and C, respectively.}
\label{fig4}
\end{figure}

To this end, we investigate the scaling connection between $\tau_\alpha$ and $S_2$ obtained at the different shear rates. Figure~\ref{fig4} shows $\tau_\alpha(t-t_d)$ as a function of $S_2(t)$ for all three samples listed in Table.1 . For sample A, data from three strain amplitudes, $\Gamma_{\text{max}}=16,\;8,\;\text{and}\;5\%$, collapse onto a single master curve. Therefore, we confirm $\tau_\alpha(t-t_d)$ is a simple monotonic function of $S_2(t)$, in non-stationary oscillatory conditions. Moreover, the collapsed data from sample A are well fit by Rosenfeld's equilibrium excess entropy scaling law,
\begin{equation}
\tau_\alpha(t-t_d)\sim e^{-cS_2(t)/k_B},
\label{eq_scaling}
\end{equation}
where $c=3.9\pm 0.2$ is a constant prefactor \cite{Rosenfeld77}.

Previous studies of the excess entropy scaling connect sample dynamics to static structure; in other words, a measurement of static structure and the scaling law can be used to predict sample dynamics. Our finding, although similar in form, has a somewhat different implication: the static structure is a consequence, rather than the cause, of the relaxation process. Slower particle rearrangement processes ($\tau_\alpha$) produce more ordered particle arrangements ($S_2$) that require longer waiting times ($t_d$) to observe. Furthermore, since the time delay, $t_d$, is explicitly encoded in Eq.~\ref{eq_scaling}, we can use information about the ``current'' sample static structure to learn about plastic flow and relaxation processes that occurred in the sample at earlier times. The structures ``remember'' sample dynamical history \cite{RevModPhys.91.035002}. In the future, application of this concept could provide insight about manufacturing and processing of amorphous materials wherein micro-structures are altered by thermomechanical processing \cite{ENGLER2002249,Sun2016}. Note also, the ``asynchronous'' dynamics-structure connection observed in our oscillatory experiments is fully compatible with steady-state experiments. When $\dot{\Gamma}$ approaches a constant value, both $\tau_\alpha(t)$ and $t_d(t)$ lose their dependence on $t$, and the new scaling framework evolves into a previous relationship found for glass-formers in uniform shear flows \cite{Tanaka18}. Thus, we expect to see this transition from a non-steady-state to a quasi-steady-state by gradually increasing the oscillatory period in future studies.

To further examine the influence of material structure and other properties on the excess entropy scaling, we plot $\tau_\alpha(t-t_d)$ versus $S_2(t)$ for samples B and C (Fig.~\ref{fig4} inset). Sample B is a monodisperse colloidal suspension (see Table.1) and thus has larger crystalline domains (see SI) compared to those in sample A. In this case, the combination of initial sample packing condition and shear-induced restructuring gives rise to much larger $|S_2|$ values, well above 4.5$k_B$, a value that corresponds to the liquid-to-crystal transition observed in two-dimensional colloidal samples \cite{Wang_JCP_2010}. By comparison, sample C is a bidisperse mixture of much smaller particles (see Table.1) and thus has more thermal particle motion; its $|S_2|$ values are between those of sample A and B. The short-time $F_s(\tau)$ for sample C is very close but never equal to unity, unlike those measured in sample A and B (see SI). We believe this difference is caused by thermal motion at short times in sample C. Despite these differences in material properties, in all three samples, both $\tau_\alpha$ and $|S_2|$ decrease with increasing shear rate (as in sample A). All experimental data thus demonstrate that excess entropy scaling with relaxation time exists independent of shear rate. The best fits to Eq.~\ref{eq_scaling} yield $c=1.0\pm 0.1$ and $1.4\pm 0.1$, for samples B and C, respectively. In previous experiments with colloidal samples, the range of the dynamics is typically one decade by utilizing multiple packing fractions. Our experiment achieves a similar dynamic range by changing shear rate alone. The excess entropy scaling form has been found to depend on factors including interfacial boundary conditions and the functional shape of the sample’s pair potentials  \cite{Ma2013,Tanaka18, doi:10.1063/1.5091564}. For our amorphous samples with small crystalline domains separated by regions of disorder, to fully understand the difference in the prefactor $c$ will require further investigation.

Finally, we also examine use of $S_2^{\theta}$ computed from directional $g(r,\theta)$; here, $\theta$ is the direction relative to shear. Unfortunately, the noise in $g(r,\theta)$ at long distances prevents Eq.~\ref{eq_entropy} from converging within the finite cut-off distance (see SI). In steady-state measurements, this sampling noise can be suppressed by time averaging. In non-steady state samples, however, time-averaging necessarily involves integration over a broader range of shear rates which complicates evaluation of sample static structure. In the future, this issue could be ameliorated by using a much larger sample size. Our observation that $\tau_\alpha$ scales with $S_2$ indicates a major difference in the microscopic relaxation mechanism between amorphous solids with small crystalline domains separated by regions of disorder and the more disordered glassy samples. In the latter, $\tau_\alpha$ has been found to scale better with the extensional excess entropy, $S_2^{\theta}$, which is derived from $g(r,\theta=\pi/4)$ \cite{PhysRevE.78.010201,Tanaka18}. The deformation along the extensional direction ($\theta=\pi/4$) has been argued to create more accessible configurations (that is, smaller $|S_2|$ values) that facilitate faster relaxation rates \cite{Tanaka18}. In sheared amorphous samples with small crystalline domains separated by regions of disorder, by contrast, particle rearrangements likely occur through cooperative sliding motions along grain boundaries, whose orientations depend on sample’s initial condition and become randomized when sample size is much larger than grain size. Therefore, we expect particle rearrangements to be less sensitive to shear in our polycrystal-like solids \cite{PhysRevE.77.042501, PhysRevE.81.051501,PhysRevLett.113.078301}.

In summary, we have developed a framework to understand plastic flow induced dynamics in deformed amorphous colloids with different degrees of polycrystallinity. The framework extends the concept of excess entropy scaling from equilibrium to nonequilibrium non-stationary states. Ours is the first experiment to demonstrate excess entropy scaling in {\it nonequilibrium} materials. Experimental data comprising a wide range of shear rates, relaxation times, particle pair correlations, and excess entropy reveal that transient shear-induced relaxation times scale as a simple exponential function of excess entropy. Collectively, these results demonstrate, in non-stationary states, that increasing (reducing) strain rates lead to faster (slower) relaxation, which in turn results in more disordered (ordered) micro-structures. The work also reveals a power-law connection between bulk shear rate and bulk viscous relaxation time that characterizes sample shear-thinning behavior; using the observation of excess entropy scaling, we thus deduce that shear-thinning is controlled by microscopic structure. Notably we find that new parameters, specifically lag times between shear rates, relaxation times and excess entropy, are crucial for proper application of the excess entropy concept in non-stationary conditions. In the future, it should be interesting to compare microscopic relaxation channels and shear-induced structural anisotropy in polycrystals versus more traditional glasses. Also, in addition to uniform and oscillatory shear, it would be desirable to test excess entropy scaling in more general strain protocols in both 2D and 3D systems.

\section{Methods}

\subsection{ Interfacial stress rheometer}
The experiments use a custom-made interfacial stress rheometer ~\cite{C3SM51014J,Reyn08}. Briefly, a pair of vertical glass walls pin a water/decane interface as shown in Fig.~\ref{fig1}a. A metal needle is located between and is parallel to the glass walls; it is held by capillary forces at the interface. Water height is adjusted so that the interface is flat between the two walls and needle. A pair of Helmholtz coils imposes a sinusoidal magnetic force on the needle that translates it axially. The out-of-plane Lorentz forces (approximately $10^{-16}$ N) are negligibly small compared with interfacial trapping forces (approximately $10^{-2}$ N). The moving needle and the two fixed boundaries thus create a flat two-dimensional (2D) shearing channel. A microscope (Infinity, K2) and high-resolution camera (IO Industries, Flare 4M180) are employed to measure the motions of the needle and interface-bound colloidal particles \cite{C3SM51014J, Shahin86}. 

\subsection{Sample preparation}
The colloidal suspensions are composed of sulfate latex particles (Invitrogen) with different diameters. The particles are injected onto the interface using a pipette, and regions of approximately 80x200 particles are studied. Due to the small particle sizes ($<10\mu$m), capillary interactions are small and unimportant \cite{PhysRevLett.45.569,KRALCHEVSKY2000145}. A long-range dipole-dipole repulsion between particles \cite{Park10}  causes the spheres to assemble into a disordered, jammed, 2D structure with large amorphous areas filling the regions between randomly oriented microcrystal domains (see Fig. \ref{fig1}b-c and SI). Characteristics of the three investigated particle systems such as particle type, packing fraction, mean interparticle separation, $d$ (derived from sample pair correlation functions), and strain amplitude are summarized in Table.1. The camera records the needle displacement and all particle motions; trajectories are extracted from the images using standard particle tracking software \cite{trackpy}. The experiments thus measure the particle positions, strain-rate, relaxation time, and excess entropy versus time during the shear cycle. We shear the samples at a fixed low frequency of 0.1 Hz to reduce/remove hydrodynamic effects. Additionally, we have calculated the Boussinesq number, $B_q=|\eta^*|/D\eta$, wherein $\eta^*$ is the complex interfacial viscosity, $D$ the needle diameter, and $\eta$ the mean viscosity of the oil and water \cite{Brooks1999,Verwijlen2011}. $B_q$ quantifies the ratio between the in-plane and out-of-plane stresses induced by the needle. We find that $B_q=147.5$ and $101.5$ for sample A and B, respectively, corroborating the expectation that hydrodynamic flows in the water and oil phases are negligible. The relaxation processes are due to plastic events that occur when the samples are stressed beyond yield.

\subsection{Affine and nonaffine particle motions}
To compute the $y$-dependent mean particle displacement, $\Delta x(y)$, along the shear ($x$) direction, we first compute $[y_j(t),\Delta x_j(t)]$ from all particles at time $t$; $\Delta x(y)$ is then obtained from the fit of $[y_j(t),\Delta x_j(t)]$ (see SI). The {\it local} strain is thus $\gamma(y,t)=\partial\Delta x(y,t)/\partial y$. To account for the slightly nonlinear flow profile  (see Fig.~S1 in SI), $\Delta x(y,t)$ and $\gamma(y,t)$ are fit by a polynomial of $y$ up to the third and second order, respectively. To characterize the overall affine deformation, we define $\Gamma(t)$ as the spatial average (over $y$) of $\gamma(y,t)$. The nonaffine particle displacement, $\Delta
\vec{r}\:'_j\equiv \{\Delta x'_j,\Delta y'_j\}$ between times $t$ and $t+\tau$ is obtained by subtracting the affine contribution from the total horizontal displacement, $\Delta x'_j(\tau)=x_j(t+\tau)-x_j(t)-\Delta x(y_j,\tau)$. Since the net flow in $y$ direction is zero, $\Delta y'_j(\tau)=y_j(t+\tau)-y_j(t)$.

\subsection{Calculation of {\bf $\langle|\dot{\Gamma}|\rangle_{\delta t}$} }
To determine $\langle|\dot{\Gamma}|\rangle_{\delta t}$, we define a triangle kernel function centered at $t$, $\Lambda(t,s)\equiv\mathrm{max}(\delta t/2-|s-t|,0)$, and we compute the convolution: $\langle|\dot{\Gamma}|\rangle_{\delta t}=(|\dot{\Gamma}|*\Lambda)(t)\equiv\int_{-\infty}^{\infty}|\dot{\Gamma(t-t')}|\Lambda(t')dt'$. (Note, $\Lambda$ is normalized before the convolution.) This parameter-free approach places maximal weight on the shear rate value at $t$, and zero weight on those shear rates outside of $[t-\delta t/2, t+\delta t/2]$. We also tested convolution with a Gaussian kernel function, and the results were very similar (see SI). 

\subsection{Evaluation of {\bf $\Delta t$} and {\bf $t_d$}}
To determine $\Delta t$, we compute the (unnormalized) correlation function, $C_1(\Delta t)\equiv\langle(\langle|\dot{\Gamma}(t)|\rangle_{\delta t}-\langle\langle|\dot{\Gamma}(t)|\rangle_{\delta t}\rangle)(\tau_\alpha^{-1}(t+\Delta t)-\langle\tau_\alpha^{-1}(t+\Delta t)\rangle)\rangle$; here, $\Delta t$ is the trial lag time and $\langle\cdots\rangle$ represents time average. The value of $\Delta t$ wherein $C_1(\Delta t)$ reaches its maximal value is set to be the true lag time between $\langle|\dot{\Gamma}(t)|\rangle_{\delta t}$ and $\tau_\alpha^{-1}(t)$ (see SI). To determine $t_d$, we similarly compute the (unnormalized) correlation function, $C_2(h)\equiv\langle|S_2(t)|\tau_\alpha(t-h\tau_\alpha)\rangle$ as a function of $h$. Similar to the procedure above, $C_2(h)$ is maximized when $h\tau_\alpha$ (or equivalently, $t_d$) is closest to the true time lag between $|S_2(t)|$ and $\tau_\alpha(t)$ (see SI).

\section{Acknowledgements}
We thank Kevin Aptowicz, Piotr Habdas, Peter Collings, Remi Dreyfus, Chandan Kumar Mishra, Alexis de la Cotte, Analisa Hill, Sophie Ettinger, Wei-shao Wei, Andrea Liu, Doug Durian, S\`{e}bastien Kosgodagan Acharige, Xiaozhou He for helpful discussions. This work was primarily supported by the National Science Foundation through Penn MRSEC Grant DMR-1720530 including its Optical Microscopy Shared Experimental Facility. Additionally, XGM and AGY gratefully acknowledge financial support from the National Science Foundation through Grant DMR16-07378, and NASA 80NSSC19K0348. KLG, DJJ and PEA gratefully acknowledge financial support from through Grant ARO W911-NF-16-1-0290.

\bibliography{references.bib}
\bibliographystyle{rsc}

\end{document}